\documentclass{PoS}
\usepackage{amsmath}
\usepackage{graphicx}
\usepackage[mathscr]{euscript}

\usepackage{physics}
\usepackage[utf8]{inputenc}

\title{Pion valence quark PDF from lattice QCD\thanks{
This work was supported by: (i) The U.S. Department of Energy, Office of Science, Office of Nuclear Physics through
the Contract No. DE-SC0012704; (iii) The U.S. Department of Energy, Office of Science, Office of Nuclear Physics
and Office of Advanced Scientific Computing Research within the framework of Sientific Discovery through Advance
Computing (SciDAC) award Computing the Properties of Matter with Leadership Computing Resources; (iii) The
Brookhaven National Laboratory's Laboratory Directed Research and Development (LDRD) project No. 16-37.
S.S. is supported by the National Science Foundation under CAREER Award PHY-1847893.
Y.Z. is also supported in part by the U.S. Department of Energy, Office of Science, Office of Nuclear Physics, within the framework of the TMD
Topical Collaboration.
The computations were carried out using USQCD facilities at JLab and
BNL under a USQCD type-A project. This research also used an award of computer time
provided by the INCITE program at the Oak Ridge Leadership Computing Facility, which
is a DOE Office of Science User Facility supported under Contract DE-AC05-00OR22725.
}
}

\ShortTitle{Pion Valence Quark PDF}

\author{
  \speaker{Charles Shugert}$^{1,2}$, Xiang Gao$^{1,3}$, Taku Izubuchi$^{1,4}$, Luchang Jin$^{4,5}$,
  Christos Kallidonis$^{2}$\footnote{Current address: Department of Physics, William \& Mary,Williamsburg, VA 23185}, 
  Nikhil Karthik$^{1}$, Swagato Mukherjee$^{1}$, Peter Petreczky$^{1}$,
  Sergey Syritsyn$^{2,4}$, Yong Zhao$^{1}$\\
\llap{$^1$}Physics Department, Brookhaven National Laboratory, Upton, NY 11973, USA \\
\llap{$^2$}Department of Physics and Astronomy, Stony Brook University, Stony Brook, NY 11794, USA
\llap{$^3$}Physics Department, Tsinghua University, Beijing 100084, China,\\
\llap{$^4$}RIKEN-BNL Research Center, Brookhaven National Lab, Upton, NY, 11973, USA \\
\llap{$^5$}Physics Department, University of Connecticut, Storrs, Connecticut 06269-3046, USA \\
  E-mail: \email{charles.shugert@stonybrook.edu}
}

\abstract{
We present lattice results on the valence-quark structure of the pion using a coordinate space method within the framework of Large Momentum
Effective Theory (LaMET). In this method one relies on the matrix elements of a Euclidean correlator in boosted hadronic states, which have an
operator product expansion at short distance that allows us to extract the moments of PDFs. We renormalize the Euclidean correlator by forming the
reduced Ioffe-time distribution (rITD), and reconstruct the second and fourth moments of the pion PDF by taking into account of QCD evolution effects.
}

\FullConference{37th International Symposium on Lattice Field Theory - Lattice2019\\
		16-22 June 2019\\
		Wuhan, China}

\begin{document}

\section{Introduction}
Parton distribution functions (PDF's) describe the longitudinal momentum distribution
of a parton within a hadron. We define the quark PDF as
\begin{equation}
  f_j(x,\mu) = \frac{1}{2}\int \frac{d\xi^-}{2\pi} e^{-ixP^+\xi^-}
  \bra{h(P)}[\overline{\psi_j}(\xi^-)\gamma^+W_L(\xi^-,0)\psi_j(0)]_{\mu}\ket{h(p)}
\end{equation}
with $P = (P^+, \frac{m_h^2}{2P^+}, 0_T)$ and $\xi = (0^+, \xi^-, 0_T)$. The parton
longitudinal momentum fraction is x and $\mu$ is the renormalization scale of the
operator. Due to the dynamical nature of the operator, it is inaccessible from direct
computation using lattice QCD. 
recently, 
the LaMET approach was proposed to extract the PDF from a Euclidean correlator in highly boosted hadronic states \cite{Ji:2013dva,Ji:2014gla}, 
which is related to
the light-cone PDF or correlator in Eq.(1) through a factorization formula \cite{Izubuchi:2018srq}. 
The factorization formula has large momentum or small distance
expansion forms in the momentum or coordinate space, respectively. The latter has been exploited in the pseudo Ioffe-time distribution approach to
extract the PDFs \cite{Radyushkin:2017cyf}, which is also used by us in this proceeding.

\subsection{Momentum space method}
The quark qPDF is defined as~\cite{Ji:2013dva}
\begin{equation}
  q_j(x,\tilde{\mu}, P_z) = \frac{1}{2}\int \frac{dz}{2\pi} e^{-ixP^z z}
  \bra{h(P)}[\overline{\psi_j}(z)\Gamma W_L(z,0)\psi_j(0)]_{\tilde{\mu}}\ket{h(p)}.
\end{equation}
Here quark-antiquark separation is in equal time, $z^\mu = (0,0,0,z)$. In addition,
we have the freedom over what gamma matrix we wish to look at. In this study, we
choose $\Gamma=\gamma_t$ to avoid operator mixing. We renormalized the qPDF matrix
element non-perturbatively in the RI-MOM scheme, then match it perturbatively to an
$\overline{\text{MS}}$-scheme PDF using the matching kernel derived from LaMET under
the form~\cite{Stewart:2017tvs}
\begin{equation}
  q_j(x,\tilde{\mu}, P_z) = \int_{-1}^{+1}\frac{dy}{|y|}
  C\big(\frac{x}{y}, \frac{yP_z}{\mu}, \frac{\tilde{\mu}}{yP_z}\big)f(y,\mu)
  + \mathcal{O}\big(\frac{m_h^2}{P_z^2}, \frac{\Lambda_{QCD}^2}{P_z^2}\big),
\end{equation}
where $\tilde \mu$ are the renormalization scales introduced in the RI-MOM scheme.
It becomes evident that large value of $P_z$ is necessary for this matching
to be reliable as it suppresses unwanted higher-twist terms. 
This approach was used to study valence PDF of the pion ~\cite{Izubuchi:2019lyk}.

\subsection{Coordinate space method}
Alternatively, one can start from the Euclidean correlator itself,
\begin{align}
\mathcal{M}(\nu, z^2; \mu^2) =
\bra{h(P)}[\overline{\psi_j(z)}\Gamma W_L(z,0)\psi_j(0)]_{\tilde{\mu}}\ket{h(p)},
\end{align}
which has also been referred to as pseudo Ioffe-time distribution \cite{Radyushkin:2017cyf}, as $\nu=z\cdot P$ is called the Ioffe-time.
We define the reduced Ioffe-Time distribution (rITD) as
\begin{equation}
  \mathscr{M}(\nu, z^2) = \mathcal{M}(\nu, z^2, \mu)/\mathcal{M}(0, z^2; \mu)
\end{equation}
The above ratio is renormalization group invariant because the renormalization
constants being $\nu$-independent cancel between the numerator and denominator.
Recent progress has been made in extracting information about the hadron PDF from
this method~\cite{Joo:2019jct,Joo:2019bzr}. In this study we extract moments of the
PDF using the following factorization formula \cite{Izubuchi:2018srq}
\begin{equation}
  \mathscr{M}(\nu, z^2) = \sum_n
  \frac{C_n(\mu^2z^2)}{C_0(\mu^2z^2)}\frac{(-i\nu)^n}{n!}a_{n+1}(\mu)
  + \mathcal{O}(z^2m_{h}^2,\; z^2\Lambda_{QCD}^2) \label{rITD_fac}
\end{equation}
where $a_{n+1}(\mu) = \int_{-1}^1 dx x^n f(x;\mu)$. Wilson Coefficients $C_n(\mu^2z^2)$
are computed in Ref.~\cite{Izubuchi:2018srq}. 
To fit lattice matrix elements computed at multiple $z^2$ values, we take into account the evolution effects in the strong
coupling and PDF moments.

\subsection{Comparison of coordinate space and momentum space methods in context of lattice calculations}
From the perspective of lattice calculations, the difference between momentum space and coordinate space method
comes from whether one fixes the momentum of the hadron to be large and
varies the separation $z$ in the quark bilinear operator, or if one varies $z$ and $P_z$ simultaneously
to cover as many values of $\nu$ as possible.
Highly boosted hadrons require special interpolating hadron operator,
tuned to optimize the overlap between with the
ground state of the hadron of a given momentum, $P_z$ \cite{Bali:2016lva}. 
In this sense, computing with the momentum space method is less expensive,
since one would only need to tune the interpolating hadron operator for
one momentum. For the coordinate space method one needs to tune the smearing operator for multiple
momenta, resulting in multiple inversions of the Dirac Operator.

An attractive feature of the coordinate space approach based on rITD is
the simplicity of the matching. There is no need to use an intermediate
RI-MOM scheme to match the lattice data to PDF defined in $\overline{\rm MS}$ scheme.
This statement, however, only holds for non-singlet case. 
In the gluon case there will be mixing with the singlet quark sector, and the matrix element at $P^z=0$ includes the moments
$\langle x_{q,g} \rangle$, which must be computed independently. 
Therefore, the gluon rITD and PDF are no longer related by a perturbative matching, as it should
include these nonperturbative moments with uncorrelated errors. 

In the momentum space method one has to perform an integral over all $z$ values
to obtain the qPDF, $\tilde q(x)$, including large $|z|$ values, where the perturbative
matching does not hold. This is not a problem for sufficiently large $P_z$ since 
the contribution from large $|z|$ is suppressed is suppressed. However, for values of $P_z$ accessible in present
day lattice calculations are no large enough to ensure this.
(see e.g. discussions in Ref.~\cite{Braun:2018brg}). In the coordinate space method
one can choose $z$ such that the perturbative matching is reliable, say $|z|<0.3$ fm. However, for
values of $P_z$ that are available in present day lattice calculations this translates to 
small values of $\nu$. As we will see in section 3 having only small $\nu$ values does not
constrain the PDF or its higher moments well. Therefore, also in the coordinate space method it is important
to consider large values of $P_z$.

\section{Details of Lattice Calculation}
In this study we calculate the valence (isovector) quark pion PDF, $f_{u}(x)-f_{d}(x)$ using
2+1 flavor gauge configurations generated by HotQCD collaboration
with the Highly Improved Staggered Quark
(HISQ) \cite{Bazavov:2014pvz}. The strange quark mass was set to its physical value, while the light quark
masses correspond to the pion mass of $161$ MeV, in the continuum limit.
We used two lattice spacings 0.06fm and 0.04fm, corresponding to $\beta=10/g^2=7.373$
and $7.825$. The
lattice volumes corresponding to coarse and fine lattices are $48^3 \times 64$ and $64^4$,
respectively. We use the Wilson-Clover action for the valence quarks on one 
HYP-smeared~\cite{Hasenfratz:2001hp} gauge background. 
We also used one HYP smeared links in gauge links that enter the spatial Wilson line.
Our valence pion mass was set to 300
MeV. For two and three point functions we incorporated All-Mode Averaging, computing 32
sloppy samples and one exact sample on each configuration~\cite{Shintani:2014vja}. Here,
sloppy and exact inversions have a stopping criteria of $10^{-4}$ and $10^{-10}$
respectively. On the coarse lattice, we have computed 100 configurations for pion
momentum 0 GeV and 0.43 GeV, and 320 configurations for momentum 0.86 GeV to 2.15 GeV.
On the fine lattice, we computed 206 configurations for all of the momentum. 
In Table \ref{tab_mom} we show the values of $P_z$ used in this study.
In order to obtain a signal for the ground state
of fast moving, 
we used Gaussian boosted sources, where a momentum $k_z$ is injected
to the quark propagator \cite{Bali:2016lva}. The values of $k_z$
are shown in Table \ref{tab_mom}.
The Gaussian source have been implemented using Coulomb gauge fixing 
\cite{Izubuchi:2019lyk} and the width of the Gaussian was set to $5.2$ in lattice
units for both lattice spacings. For the coarse lattices our setup is identical
to the one used in Ref. \cite{Izubuchi:2019lyk}.
\begin{table}
  \centering
  \begin{tabular}{| c | c | c | c | c | c | c |}
    \hline
    $n_z$            & 0 & 1    & 2    & 3    & 4   &  5   \\ \hline
    $P_z$ GeV coarse & 0 & 0.43 & 0.86 & 1.29 & 1.72 & 2.15 \\ \hline
    $P_z$ GeV fine   & 0 & 0.48 & 0.97 & 1.45 & 1.93 & 2.42 \\ \hline
    $\zeta$          & 0 & 0    & 1    & 0.66 & 0.75 & 0.6 \\
    \hline
  \end{tabular}
  \caption{
    The pion momenta, $P_z=2 \pi/L n_z$ and the 
    boost parameter $\zeta=k_z/P_z$ used 
    in the valence quark propagator.}
\label{tab_mom}
\vspace*{-0.4cm}
\end{table}
We extract ground state matrix elements using the summation
method \cite{Maiani:1987by}.

\section{Constraints on pion PDF from rITD}
\subsection{Studying rITD using Phenomenological Results}
\begin{figure}
  \centering
  \begin{minipage}{0.5\textwidth}
    \centering
    \includegraphics[width=0.95\textwidth]{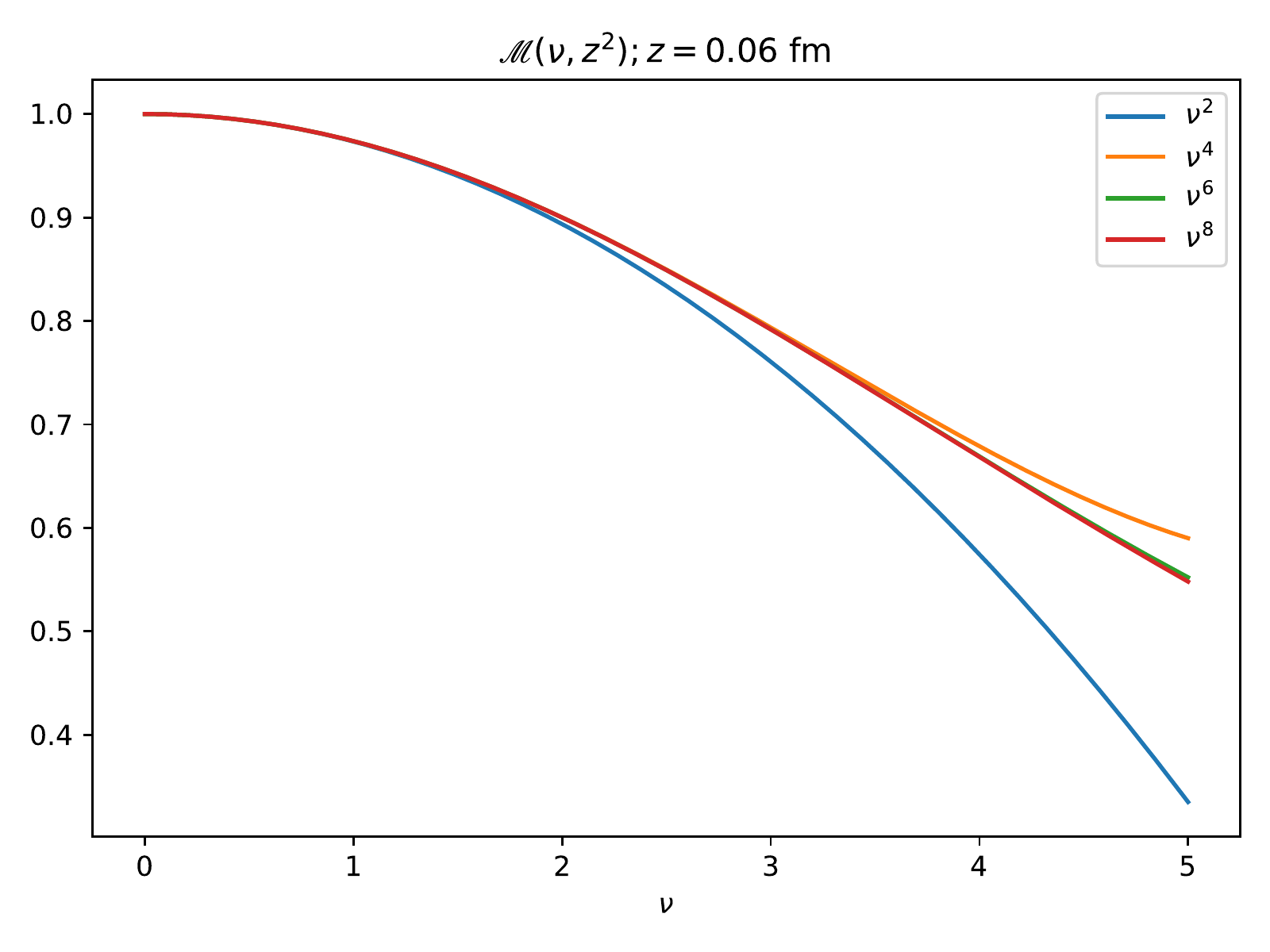}
    \caption{
      Truncated rITD from JAM result. \newline
      Different curves follow different truncation in $\nu$. \label{fig:rITD_jam_trunc}
    }
  \end{minipage}\hfill
  \begin{minipage}{0.5\textwidth}
    \centering
    \includegraphics[width=0.95\textwidth]{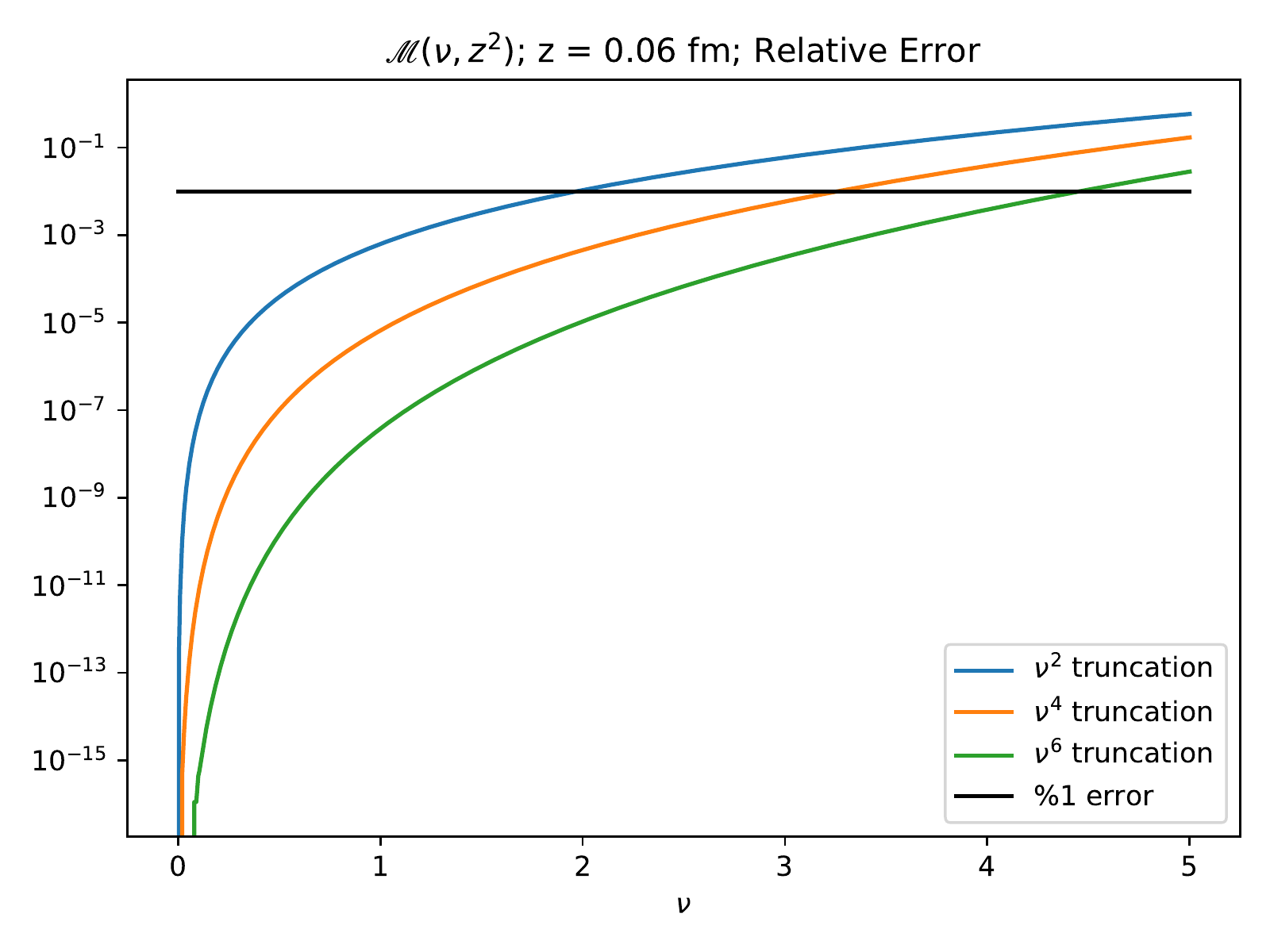}
    \caption{
      Relative error between different truncation's in $\nu$ of rITD from JAM result.
      The black curve shows 1 $\%$ relative error. \label{fig:rITD_jam_err}
    }
  \end{minipage}\hfill
\end{figure}

To demonstrate how the rITD can constrain
the valence quark pion PDF we use the results from phenomenological
analysis of Drell-Yan data by JAM collaboration \cite{Barry:2018ort}.
First we notice that in the isospin symmetric limit, the valence quark
distribution is symmetric around $x=0$ (see discussions in Ref. \cite{Izubuchi:2019lyk}). 
Therefore, all the odd
moments of the valence quark PDF will vanish.
The JAM result for the pion valence quark PDF at $\mu=3.2$ GeV and $x>0$
can be parameterized using a simple
form $f(x) \sim x^a(1-x)^b$, $a=-0.407$ and $b = 1.12$. 
Using this parameterization of PDF it is straightforward to calculate the moments
and reconstruct the rITD corresponding to JAM result according to Eq. (\ref{rITD_fac}).
Since in the lattice calculations we can only access a limited range of Ioffe-time $\nu$
the series in Eq. (\ref{rITD_fac}) can be truncated to relatively low order. This is shown in
Figs. \ref{fig:rITD_jam_trunc} and \ref{fig:rITD_jam_err}.
If $\nu$ is smaller than 5  rITD can be described by 8th order polynomial. This implies
that present day lattice calculations cannot constrain moments beyond $\langle x^8 \rangle$.
From the above figures  we can also see that $\langle x^4 \rangle$ terms begin to contribute
at $\nu \sim 2$ and $\langle x^6 \rangle$ terms begin to contribute at $\nu \sim 3$.

\begin{figure}
\includegraphics[width=0.45\textwidth]{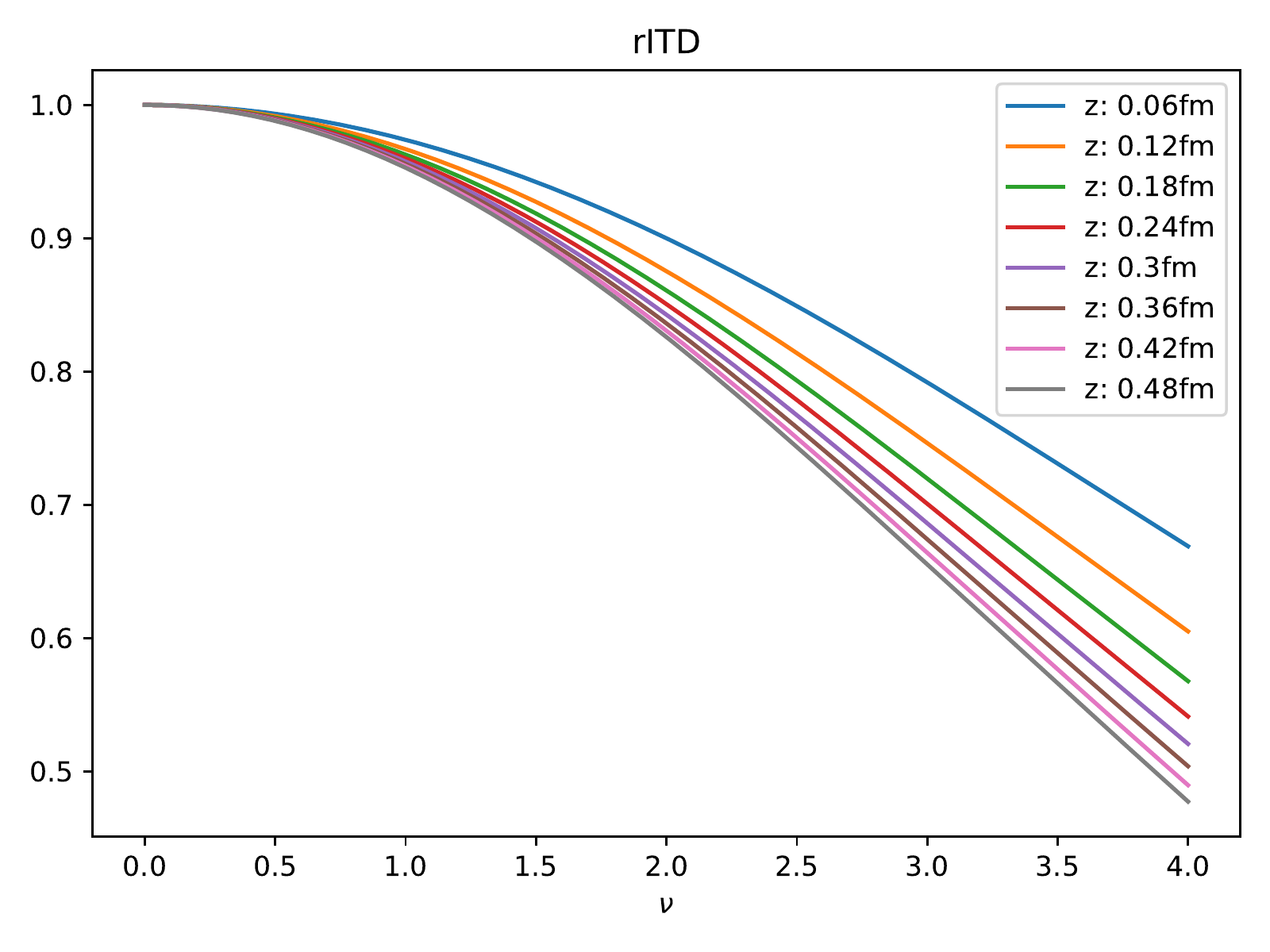}
\includegraphics[width=0.45\textwidth]{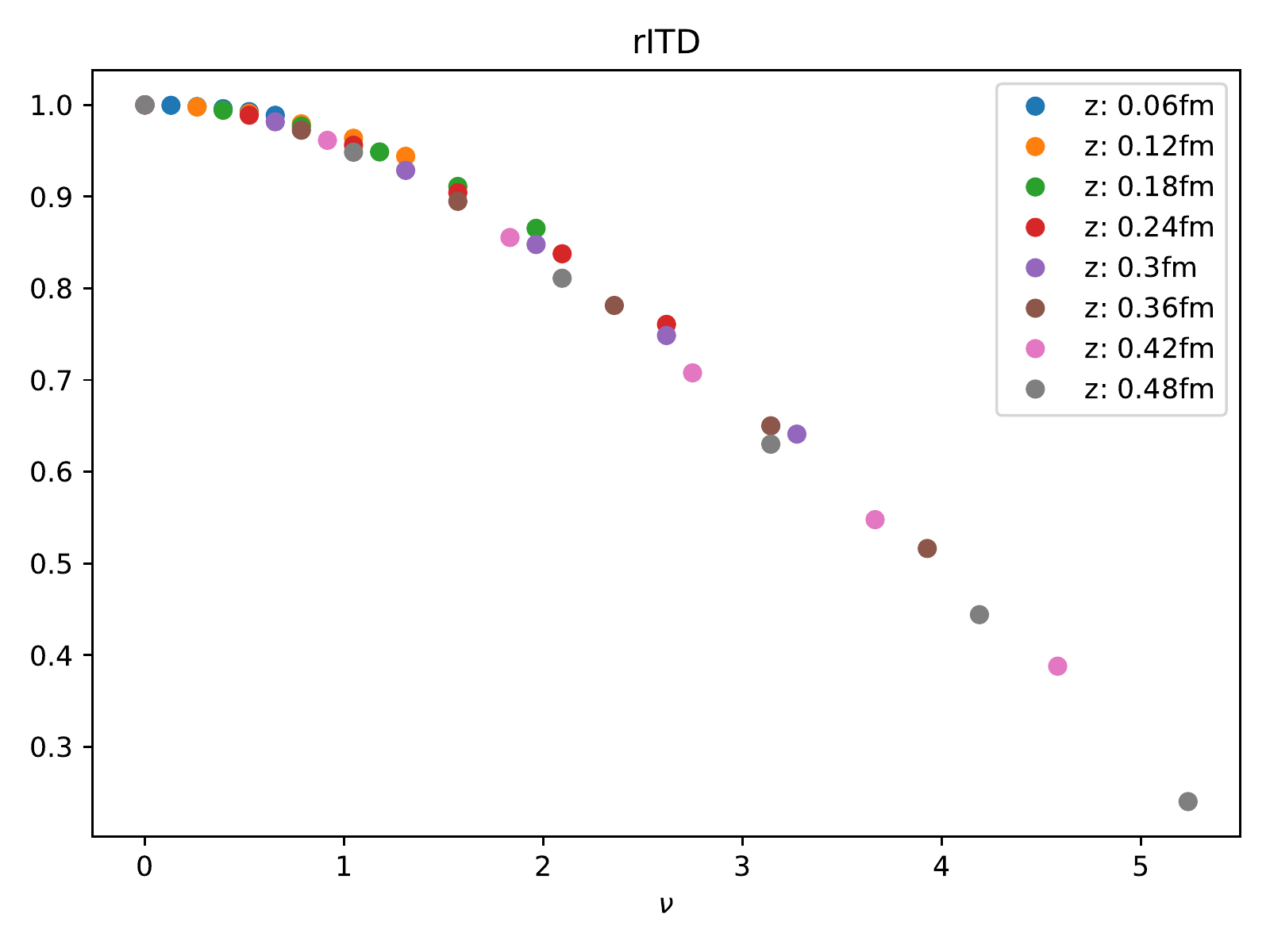}
\caption{
The $z^2$ dependence of 
Ioffe time for $\nu \in [0, 5]$ obtained from JAM result on valence quark
pion PDF. The left panel shows the general case, while the right panel shows
the $z$-dependence for the set of the $z-$ values and momenta, $P_z$ available
in lattice calculations with $a=0.06$ fm, c.f. Table 1.
}
\label{fig:rITD_zdep}
\end{figure}

Next we explore the $z^2$ evolution of the rITD
using the JAM results for the valence pion PDF 
This is shown in Fig. \ref{fig:rITD_zdep}
From this figure we see that
at large Ioffe time the $z^2$ dependence of rITD could be significant. However,
in lattice calculations only certain values of $\nu$ can be probed because
both $P_z$ and $z$ can be varied in certain steps. In the right panel of Fig. \ref{fig:rITD_zdep}
we show then $z^2$ dependence of rITD for our lattice setup corresponding to $a=0.06$ fm.
It is evident from the figures that for our lattice setup the effects of $z^2$ evolution are small.

\subsection{rITD from Lattice Data}
\begin{figure}
  \centering
  \includegraphics[scale=0.5]{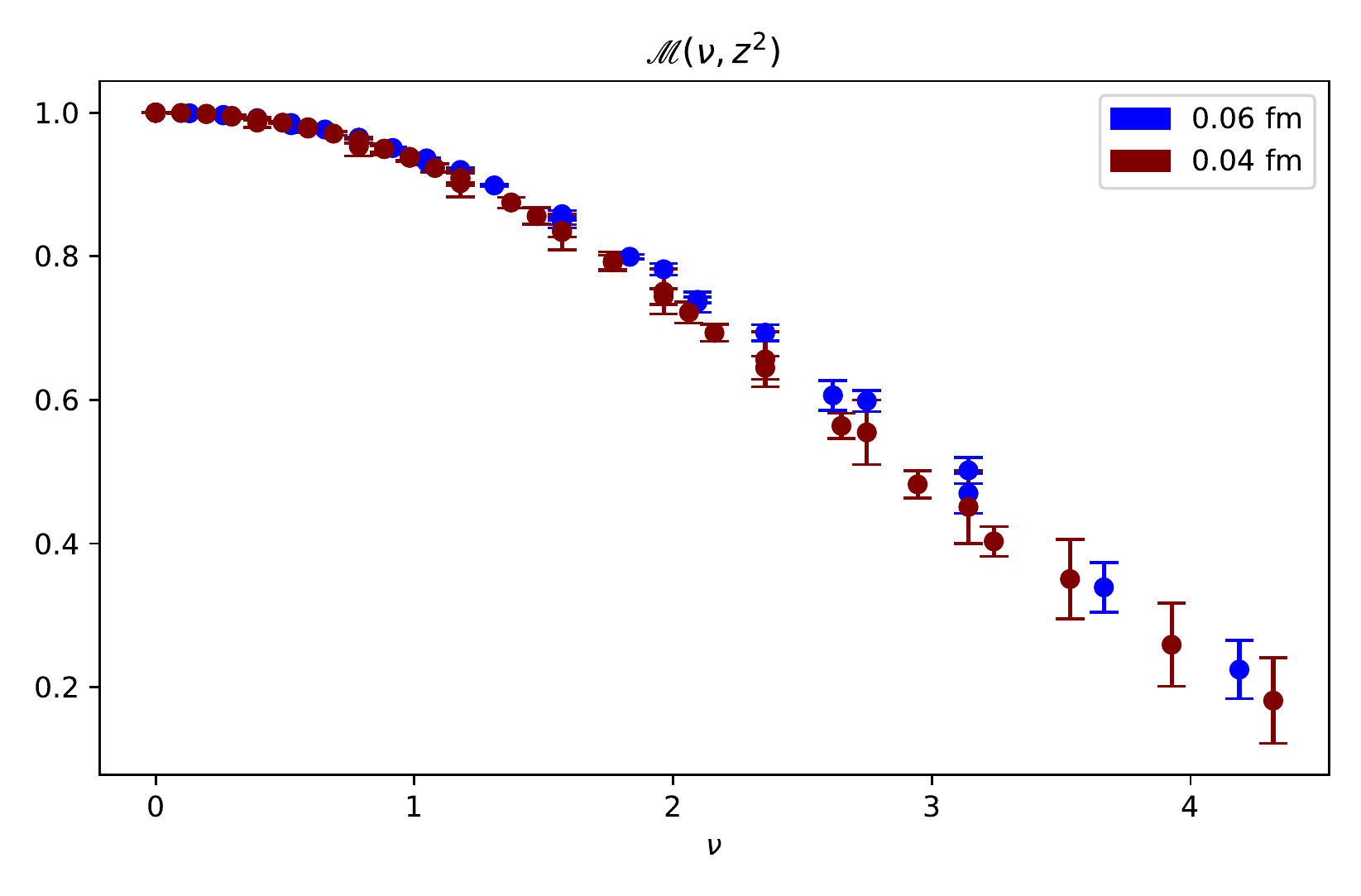}
  \caption{
    Lattice results for rITD for $a=$0.04 fm and 0.06 fm corresponding 
    to $z \le 0.48$ fm. The 
    largest momenta used are 1.93 GeV and 1.72 GeV for $a=0.04$ fm and $a=0.06$ fm, respectively. 
  }
\label{fig:rITD_lat}
\end{figure}

We calculated the valence quark rITD of the pion for two lattice spacings, $a=0.06$fm and $a=0.04$ fm.
To reduce the errors of the lattice calculations we consider the following 
quantity
$$  \mathscr{M}^{\text{imp}}(\nu, z^2) =
  \frac{\mathcal{M}(\nu, z^2)}{V(P)}\frac{V(0)}{\mathcal{M}(0, z^2)},$$
with $V(P) = \bra{P}\bar{\psi}(0)\gamma_0\psi\ket{P}$ as suggested
in Ref. ~\cite{Orginos:2017kos,Karpie:2017bzm}. Since the renormalization of the electric charge does not
depend on the hadron momentum, multiplying the rITD by $V(0)$ and dividing by $V(P_z)$ does not change anything.
However, the matrix elements calculated for the same hadron momentum are strongly correlated and therefore,
using the above ratio strongly reduces the errors for the lattice estimate of rITD.
The corresponding estimate for rITD are shown in Fig. \ref{fig:rITD_lat}
This figure shows that rITD calculated at two lattice spacings agree within errors, and thus
cutoff effects are small even at small values of $z$. The $z$-dependence of rITD at fixed $\nu$ 
is small for the considered range of $z$ values in agreement with the expectations based on
the analysis of rITD obtained from the JAM valence quark pion PDF.

\begin{figure}
  \centering
  \begin{minipage}{0.4\textwidth}
    \centering
    \includegraphics[width=0.95\textwidth]{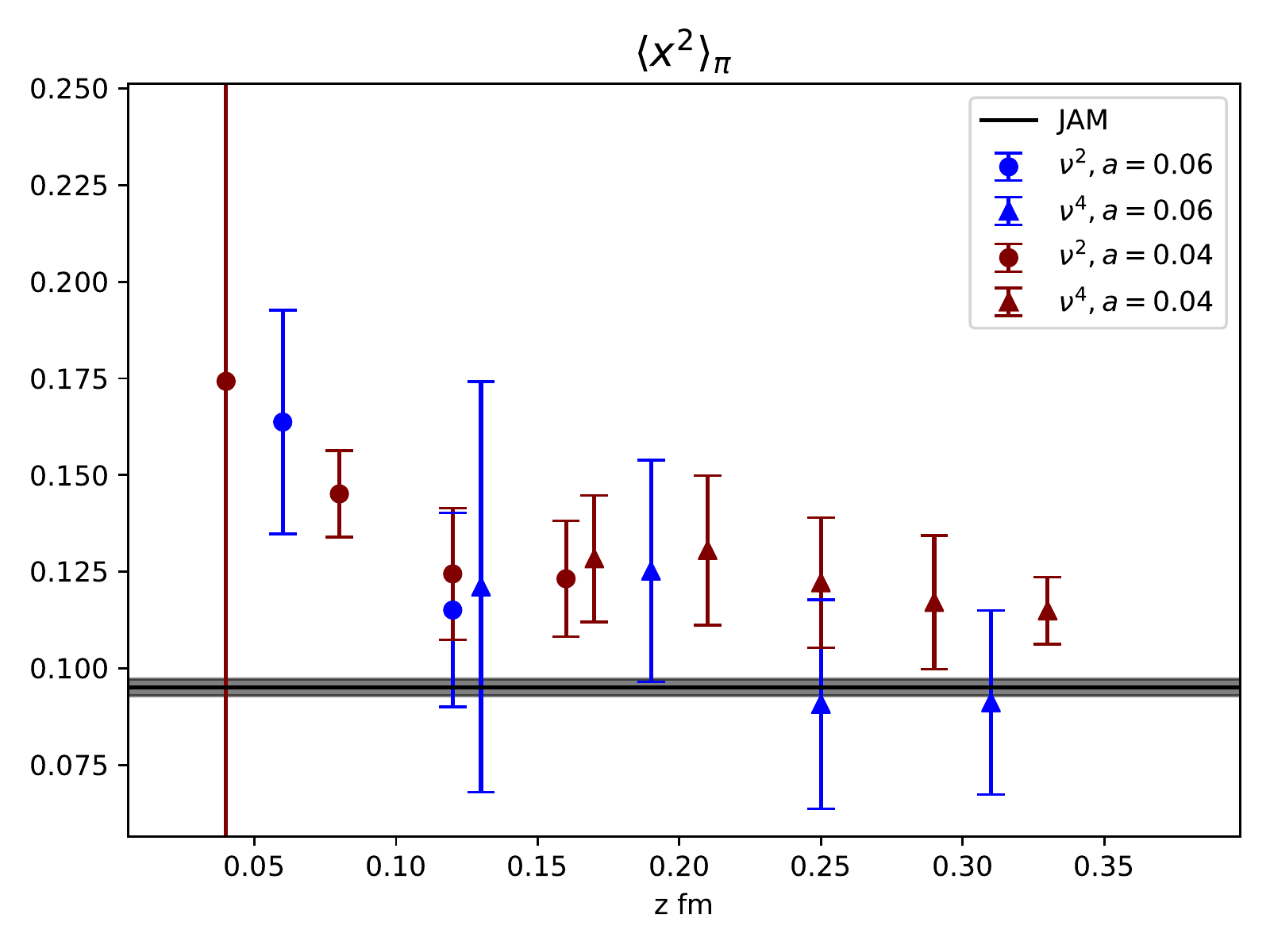}
    \caption{
      Fit to $\langle x^2 \rangle$. Circle points are from a quadratic fit. Triangular
      points are from a quartic fit. Blue points are fits using the coarse lattice.
      Maroon points are fits using the fine lattice.
    }
    \label{fig:mom2}
  \end{minipage}\hfill
  \begin{minipage}{0.4\textwidth}
    \centering
    \includegraphics[width=0.95\textwidth]{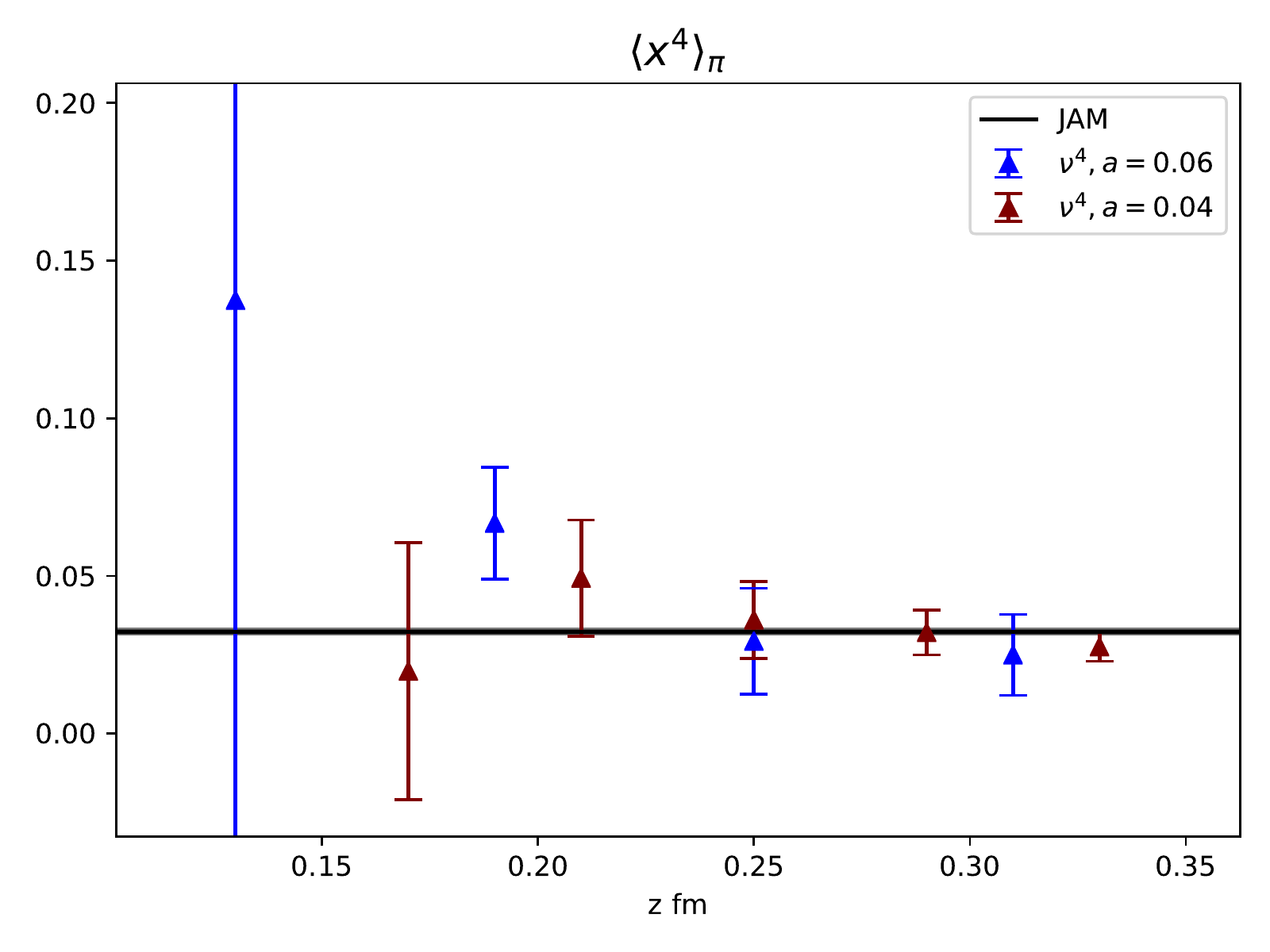}
    \caption{
      Fit to $\langle x^4 \rangle$. Blue points are fits using the coarse lattice.
      Maroon points are fits using the fine lattice.
    }
  \label{fig:mom4}
  \end{minipage}\hfill
\end{figure}

To determine the moments of the PDF, we perform a combined fit across all of the
lattice data up to some $z_o$ to a truncated polynomial in $\nu$ of the
form~\eqref{rITD_fac} at renormalization scale $\mu=3.2$ GeV, shown in Figs \ref{fig:mom2} and \ref{fig:mom4}.
Currently, more statistics is  needed to quote a value for the moments. However, one
can see that the fit appears to stabilize if one includes larger $z^2$ values. This
however introduces systematic uncertainties coming from non-perturbative physics
and target-mass effects. 
We will try to estimate these effects in the future.

\section{Conclusions}
We have presented results of pion PDF using the OPE of a Euclidean correlator in boosted pion states.
In addition we studied the $z^2$ evolution of the rITD as well as the behavior
due to truncation as a function of $\nu$. We find large dependence in $z^2$. We also
find how truncated fits are sensitive to the Ioffe-Time available. We present our rITD
at two different lattice spacings. Finally we fit our rITD to quadratic and quartic
polynomials in order to extract the second and fourth moments of valence quark pion PDF.



\bibliographystyle{JHEP}
\bibliography{my_bib.bib}


\end{document}